\documentclass[a4paper,11pt]{article}
\pdfoutput=0

\usepackage{jcappub}
\usepackage{natbib}
\usepackage{dcolumn}
\usepackage{graphicx}
\usepackage{color}
\usepackage{bm}
\usepackage{bbold}
\usepackage[usenames,dvipsnames]{xcolor}
\usepackage[normalem]{ulem}
\usepackage{multirow}
\usepackage{ulem}

\newcommand{\apj}{Astrophys.~J.}
\newcommand{\mnras}{Mon.~Not.~R.~Astron.~Soc.}

\newcommand{\be}{\begin{equation}}
\newcommand{\ee}{\end{equation}}
\newcommand{\bea}{\begin{eqnarray}}
\newcommand{\eea}{\end{eqnarray}}

\newcommand{\E}{\mathcal{E}}

\title{The effects of structure anisotropy on lensing observables in an exact general relativistic setting for precision cosmology}

\author[a]{M. A. Troxel}
\author[a]{Mustapha Ishak}
\author[a]{Austin Peel}

\affiliation[a]{Department of Physics, The University of Texas at Dallas, \\Richardson, TX 75080, USA}

\abstract{The study of relativistic, higher order, and nonlinear effects has become necessary in recent years in the pursuit of precision cosmology. We develop and apply here a framework to study gravitational lensing in exact models in general relativity that are not restricted to homogeneity and isotropy, and where full nonlinearity and relativistic effects are thus naturally included. We apply the framework to a specific, anisotropic galaxy cluster model which is based on a modified NFW halo density profile and described by the Szekeres metric. We examine the effects of increasing levels of anisotropy in the galaxy cluster on lensing observables like the convergence and shear for various lensing geometries, finding a strong nonlinear response in both the convergence and shear for rays passing through anisotropic regions of the cluster. Deviation from the expected values in a spherically symmetric structure are asymmetric with respect to path direction and thus will persist as a statistical effect when averaged over some ensemble of such clusters. The resulting relative difference in various geometries can be as large as approximately $2\%$, $8\%$, and $24\%$ in the measure of convergence ($1-\kappa$) for levels of anisotropy of $5\%$, $10\%$, and $15\%$, respectively, as a fraction of total cluster mass. For the total magnitude of shear, the relative difference can grow near the center of the structure to be as large as $15\%$, $32\%$, and $44\%$ for the same levels of anisotropy, averaged over the two extreme geometries. The convergence is impacted most strongly for rays which pass in directions along the axis of maximum dipole anisotropy in the structure, while the shear is most strongly impacted for rays which pass in directions orthogonal to this axis, as expected. The rich features found in the lensing signal due to anisotropic substructure are nearly entirely lost when one treats the cluster in the traditional FLRW lensing framework. These effects due to anisotropic structures are thus likely to impact lensing measurements and must be fully examined in an era of precision cosmology.}

\begin{document}

\nocite{*}

\maketitle

\section{Introduction}\label{intro}

As the quality of astrophysical data rapidly improves over the next decades, we are presented with rich opportunities to consider and constrain models of both large- and small-scale structure in the universe which take into account the true, inhomogeneous and anisotropic nature of structure. With goals to constrain cosmology at the percent level utilizing upcoming survey results, the consideration of second-order or nonlinear effects is becoming a priority \cite{bernardeau1,bernardeau2,jeong,ido1,umeh,ido2,ido3,challinor}. One such avenue of exploration is in employing structure models, which are exact solutions to Einstein's field equations in general relativity but are not restricted to homogeneity and isotropy, and developing the methods necessary to compare observational predictions of these models to new and improving data \cite{SKMHH2003,Krasinski1997}. This includes the identification of what impact inhomogeneities and anisotropies will have on astrophysical observables such as lensing and dynamical mass estimates. While this is worthwhile in its own right as an examination of what more general exact models of structure in general relativity predict, it also informs us of biases in the interpretation of observational data which symmetry assumptions might cause. These models also take into account the full nonlinearity of general relativity without assumptions or simplifications, so that higher order and relativistic effects which have recently attracted attention in the literature are automatically included.

One of the most promising probes of the universe is gravitational lensing, either in the form of strong and weak lensing by galaxies and clusters of galaxies or of weak lensing by large-scale structure in the universe (cosmic shear). The combination of new lensing measurements from large ongoing and planned surveys with other probes like the cosmic microwave background and the distance-redshift relation from type Ia supernovae promises to significantly constrain cosmological information. Including cosmic shear, for example, can improve constraints of cosmological parameters by factors of 2-4 (e.g. \cite{tj}). Developing a framework in which we can identify and investigate observables like the lensing convergence and shear within these general inhomogeneous and anisotropic exact solutions is oftentimes not straightforward, and has been the topic of much recent work. We continue this here by describing a framework in which to study gravitational lensing in such general metrics, using the Szekeres metric as an example, and discuss how it can be associated with the traditional lensing convergence and shear used in studies of the universe in the concordance Lambda cold dark matter ($\Lambda$CDM) model, which is based upon the Freidmann-Lema\^itre-Robertson-Walker (FLRW) metric. The lensing effects of anisotropy on cluster scales have primarily been studied recently in terms of the potential triaxiality of a dark matter halo, where determinations of mass through gravitational lensing have been shown to be affected by up to 50\% for cases where the halo is significantly elongated along the line of sight \cite{Corless2007}. In weak lensing mock catalogues, determinations of mass and concentration can be impacted at the 5\% level \cite{Bahe2012}. Inhomogeneous structure has also been used to demonstrate potential challenges to precision cosmological constraints, for example, due to lensing of very small beam sources like supernovae \cite{clarkson}. 

The Szekeres metric \cite{Szekeres1,Szekeres2} is an exact solution which has been used in cosmology, but is also able to model realistic, small-scale structures with a natural FLRW limit as background. The Szekeres models have an irrotational dust source with no symmetries (i.e., no Killing vectors \cite{Bonnoretal1977}), and thus can represent structure with no assumptions of spherical or axial symmetry. Previous attempts to compare Szekeres models to cosmological observations have included the growth of large scale structure \cite{Ishak&Peel2012,Peel2012}, the expansion history of the universe \cite{Ishaketal2008,Bolejko&Celerier2010,Nwankwoetal2011}, as well as some cosmic microwave background constraints \cite{BolejkoCMB,Buckley}. On smaller scale, they have been studied as models of exact structures like clusters of galaxies, primarily as a means to test whether realistic cluster-sized densities can evolve in a Szekeres universe from reasonable initial conditions and without singularities \cite{bolejko2007,krasinski2002,hellaby2006,walters2012}. In the past, it has been shown that structure growth on both small and large scales is enhanced in the nonlinear Szekeres models relative to $\Lambda$CDM \cite{Bolejko2006,Ishak&Peel2012,Peel2012,troxel1}. A systematic consideration of gravitational lensing in the Szekeres metric has not yet been attempted, though some initial work has been done for the LT metric \cite{ghassemi2009,masina2009,parsimood2013a,parsimood2013b}. {\cite{bolejkovoid}, for example, have considered the impacts of voids (using Szekeres models) on the measured magnitude of astronomical objects such as supernovae type Ia, while \cite{meures} have commented on the cosmological constant and lensing in class II Szekeres models.

This paper follows as part of a series which seeks to explore the effects of inhomogeneity and anisotropy on astrophysical observables in exact, relativistic models, using the Szekeres models as a test case. In an era of precision cosmology, such effects have been demonstrated to be significant (e.g. \cite{Peel2012,troxel1} and references). The Szekeres models possess both the homogeneous FLRW models and the spherically symmetric Lema\^itre-Tolman (LT) models \cite{Lemaitre1933,tolman1934} as natural limits, which aides in the systematic exploration of the effects due to inclusion of anisotropy in structures. In \cite{troxel1}, we developed a realistic Szekeres cluster model, which reproduces the density profile of a cluster of galaxies at $t_0$ (the current age of the universe) based upon the Navarro-Frenk-White dark matter profile \cite{nfw97}. We showed that the inclusion of anisotropy relative to a reference spherically symmetric LT model produces a strong, nonlinear response in the rate of gravitational clustering in anisotropic regions of the structure, which contributes to a greater total infall velocity. Here we develop a framework in which to calculate the lensing properties of such a Szekeres cluster model with varying levels of anisotropy by examining the deviation of neighboring geodesics as they propagate through and past the structure. This is a first step to a more general examination of gravitational lensing in the Szekeres metric.

The paper is structured as follows. In section \ref{lensing}, we introduce the lensing framework for inhomogeneous cosmologies, which has been derived from a general treatment of geometric optics in general relativity. Section \ref{relating} then describes how this general treatment can be related to the classic lensing formalism in the case that there exists a background FLRW cosmology, and we derive the convergence and shear for our particular cluster density model in section \ref{lcdmcluster}. In section \ref{modeldef} we discuss the calculation of general lensing properties in an exact, anisotropic model. The Szekeres metric is introduced in section \ref{anisotropic}, where we briefly describe how anisotropies can be systematically incorporated into the model, and we specialize the general lensing formalism to calculations in the Szekeres metric in section \ref{geodev}. As an example of the process, we examine what effects the introduction of anisotropies in a structure have on its lensing properties in section \ref{anidev}. Finally, we conclude in section \ref{conclusion}. Units are chosen throughout the paper such that $c=G=1$.

\section{Gravitational lensing in inhomogeneous cosmologies}\label{lensing}
\subsection{General framework}\label{framework}

The framework for the exact study of lensing in general inhomogeneous cosmologies differs substantially from the traditional lensing framework in the Lambda Cold Dark Matter ($\Lambda$CDM) paradigm, where several simplifying assumptions can safely be made (see for example \cite{SEF,seitz} and references therein). In a general curved spacetime, even the electromagnetic wave equation obtains a first-order Ricci tensor component, 
\be
\square A^{\mu}=-\mu_0J^{\mu}+R^{\mu}{}_{\nu}A^{\nu},
\ee
which causes the self-interaction of the wave propagation with the curvature it produces in the spacetime. Thus the treatment of a light ray's propagation as geodesic and affinely parameterized is only possible as a zeroth order approximation. This approximation is satisfied in the short-wave (WKB) limit, which we employ here, such that to an observer the electromagnetic waves appear sufficiently planar and monochromatic over large scales compared to a typical wavelength, while remaining small compared to the radius of curvature of the space. In this limit, we can ignore the contribution of the curvature to the wave equation. However, it is important to keep this initial assumption in mind, as it impacts in an exact treatment even the well-known approximation for the redshift of light, where in the FLRW metric, for example, the redshift can be related to the scale factor $a$ by
\be
1+z=\frac{k^{\mu}u_{\mu}|_{e}}{k^{\mu}u_{\mu}|_{o}}=\frac{a_0}{a},
\ee
where $k^{\mu}$ is the null tangent vector of the light and $u^{\mu}$ is the 4-velocity of the observer.

In order to study gravitational lensing by mass in a general cosmology, we will consider the geodesic deviation of adjacent light rays $\gamma$ in a bundle, which will be parameterized relative to some fiducial ray $\gamma_0$. This treatment follows the work of \cite{seitz}, and we will use their notation where possible. It is described in terms of our specific use, but the reader is directed to \cite{seitz} for a thorough treatment of the formalism. An alternative approach using the geodesic light-cone gauge has also recently been discussed in \cite{fanizza}, where some of the following equations can be given explicit solutions. The fiducial ray will be described by some null tangent vector $k^{\mu}=\partial x^{\mu}/\partial \lambda$, which is affinely parameterized by a parameter $\lambda$. For a past-directed light cone, $\lambda=0$ at the observer and increases going backward in time. At the observer, the following is then true for a metric signature $(-+++)$:
\bea
(u^{\mu}u_{\mu})_0&=&-1\\
u_0^{\mu}k_{\mu}&=&1.
\eea

The bundle of light (taken to represent, for example, the image of a distant galaxy) is then propagated along the path of the fiducial ray, and the deviation of nearby rays relative to the fiducial ray describes the lensing properties of any intervening mass. To do this, we define a `deviation vector field' or Jacobi field
\be
Y^{\mu}(\vec{\theta},\lambda)=\gamma^{\mu}(\theta,\lambda)-\gamma_0^{\mu}(\theta=0,\lambda),
\ee
such that $Y$ represents the (changing) separation between some ray $\gamma$ and the fiducial ray, from which the angular position $\theta$ is identified. $Y$ is parameterized by a space-like basis $(E_1,E_2)$ which span the plane orthogonal to both $u^{\mu},k^{\mu}$ to complete the tetrad along $\gamma_0$, where $u^{\mu}$ is the vector resulting from parallely propagating $u^{\mu}_0$ along $\gamma_0$. We can then rewrite $Y$ as
\be
Y^{\mu}=\xi_1 E_1^{\mu}+\xi_2 E_2^{\mu}+\xi_0 k^{\mu}.
\ee

The components of $Y$ along this orthogonal screen ($\xi_1,\xi_2$) can then be described by the deformation equation, which in matrix form is
\be
\dot{\boldsymbol{\xi}}=\boldsymbol{\mathcal{S}}\boldsymbol{\xi},\label{eq:deformation}
\ee
where at each point $\lambda$ along the fiducial ray
\be
\boldsymbol{\mathcal{S}}=\begin{pmatrix} \theta-\Re \sigma&\Im \sigma\\
\Im \sigma&\theta+\Re \sigma\\\end{pmatrix}
\ee
is the optical deformation matrix, composed of the Sachs optical scalars \cite{sachs}: the rate of expansion,
\be
\theta=\frac{1}{2}k^{\mu}{}_{;\mu},
\ee
and the complex rate of shear,
\be
\sigma=\frac{1}{2}k_{\mu;\nu}\epsilon^{*\mu}\epsilon^{*\nu},
\ee
where $\epsilon^{\mu}=E_1^{\mu}+i E_2^{\mu}$. 

Taking the derivative of equation (\ref{eq:deformation}) w.r.t. $\lambda$ gives a more useful description of the evolution of ($\xi_1,\xi_2$) along the fiducial ray, however, which can be written in terms of the Ricci and Weyl curvatures of the spacetime. We can then write 
\be
\ddot{\boldsymbol{\xi}}=\boldsymbol{\mathcal{T}}\boldsymbol{\xi},\label{eq:tidaleq}
\ee
where the optical tidal matrix $\boldsymbol{\mathcal{T}}$ is given by
\be
\boldsymbol{\mathcal{T}}=\begin{pmatrix} \mathcal{R}-\Re \mathcal{F}&\Im \mathcal{F}\\
\Im \mathcal{F}&\mathcal{R}+\Re \mathcal{F}\\\end{pmatrix}.
\ee
The source of convergence is written in terms of the Ricci curvature,
\be
\mathcal{R}=-\frac{1}{2}R_{\mu\nu}k^{\mu}k^{\nu},
\ee
while the source of shear is written in terms of the Weyl curvature,
\be
\mathcal{F}=-\frac{1}{2}C_{\alpha\beta\mu\nu}\epsilon^{*\alpha}k^{\beta}\epsilon^{*\mu}k^{\nu}.
\ee
Once one has defined a spacetime, the Ricci and Weyl curvatures can then be calculated, and equation (\ref{eq:tidaleq}) describes the deformation (or lensing) of such an infinitesimal light bundle as it propagates as $\xi_i$ on the screen at each event parameterized by $\lambda$ along the fiducial ray.

\subsection{The geodesic deviation and the standard lensing framework in perturbed $\Lambda$CDM}\label{relating}

One can use a Jacobi mapping between $\xi_i$ and the initial angle $\theta_i$ between the ray of interest and fiducial ray, $\boldsymbol{\xi}(\lambda)=\boldsymbol{\mathcal{D}}(\lambda)\boldsymbol{\theta}$, due to the linearity of equation (\ref{eq:tidaleq}). The evolution of $\boldsymbol{\mathcal{D}}$ as a function of $\lambda$ is then given by
\be
\ddot{\boldsymbol{\mathcal{D}}}=\boldsymbol{\mathcal{T}}\boldsymbol{\mathcal{D}},\label{eq:tidaleq2}
\ee
where at $\lambda=0$, $\boldsymbol{\mathcal{D}}=\mathbb{0}$ and $\boldsymbol{\dot{\mathcal{D}}}=\mathbb{1}$. This Jacobi matrix $\boldsymbol{\mathcal{D}}$ is related to the traditional magnification matrix $\boldsymbol{\mathcal{A}}$ of lens theory when properly scaled. 

In an FLRW model, there exists no Weyl curvature and so $\boldsymbol{\mathcal{T}}=\mathcal{R}\mathbb{1}$, which implies that $\boldsymbol{\mathcal{D}}=\tilde{\mathcal{D}}\mathbb{1}$ and $\tilde{\mathcal{D}}$ becomes the angular diameter distance $D_A$. In a general space, one finds instead that $D_A=\sqrt{\det{\boldsymbol{\mathcal{D}}}}$. From the field equations, it can be shown that in FLRW the source of convergence simplifies to
\be
\mathcal{R}=-4\pi\rho_{bg}(1+z)^2=-\frac{3}{2}H_0^2\Omega_m^0(1+z)^5.
\ee
In the weak field limit, we can also include an explicit, isolated mass density (e.g. a cluster of galaxies) and express the optical tidal matrix directly as a function of the density (or alternately the Newtonian or lensing potential). In this case, the source of convergence becomes simply the sum of these two masses
\be
\mathcal{R}=-4\pi(\rho_{bg}+\rho_{cl})(1+z)^2.
\ee
Over large distances, the cluster density can be neglected, as it acts only over a very small part of the total null path of the rays. The source of shear can be written in terms of the Newtonian potential as
\be
\mathcal{F}=-(2\Phi,_{ij}+\delta_{ij}\Phi,_{33})(1+z)^2.
\ee

Writing the optical tidal matrix in the FLRW case as above is one method of comparing results between the standard lens theory and the resulting information on geodesic deviation which comes from the solution of equation (\ref{eq:tidaleq2}). Another method is to instead relate the lensing convergence ($\kappa$) and shear ($\gamma$) to the Jacobi matrix $\boldsymbol{\mathcal{D}}$ as mentioned above. It is straightforward to show that when we consider an inhomogenous model with a matching or limiting background homogeneous cosmology, this relationship is given by
\be
\boldsymbol{\mathcal{D}}=\tilde{D}_A\boldsymbol{\mathcal{A}},
\ee
where $\tilde{D}_A$ is the angular diameter distance in the background model and $\boldsymbol{\mathcal{A}}$ is 
\be
\boldsymbol{\mathcal{A}}=\begin{pmatrix} 1-\kappa-\gamma_1&\gamma_2\\
\gamma_2&1-\kappa+\gamma_1\\\end{pmatrix}.
\ee
Thus given a solution to equation (\ref{eq:tidaleq2})
\be
\boldsymbol{\mathcal{D}}=\begin{pmatrix} D_{11}&D_{12}\\
D_{21}&D_{22}\\\end{pmatrix}
\ee
and assuming a general model which can be decomposed into a discrete mass structure (or structures) with an associated homogeneous background cosmology, we can then write the convergence and shear due to a structure as (e.g. \cite{clarkson})
\bea
\kappa&=&1-\frac{D_{11}+D_{22}}{2\tilde{D}_A}\\
\gamma_1&=&\frac{D_{22}-D_{11}}{2\tilde{D}_A}\\
\gamma_2&=&\frac{D_{12}}{\tilde{D}_A}=\frac{D_{21}}{\tilde{D}_A}\\
\gamma&=&\sqrt{\gamma_1^2+\gamma_2^2}.\label{eq:dtokappa}
\eea
In a purely FLRW model where $\boldsymbol{\mathcal{A}}=\mathbb{1}$, this result is consistent with the above assertion that $\boldsymbol{\mathcal{D}}=\tilde{D}_A\mathbb{1}$, with $\tilde{D}_A$ just being the angular diameter distance in the model.

\subsection{Lensing due to a modified NFW galaxy cluster in a $\Lambda$CDM background}\label{lcdmcluster}

In \cite{troxel1}, we explored the kinematic impact of introducing anisotropies into the density profile of a cluster of galaxies. We chose the truncated NFW density model of \cite{bmo}, but for reasons of regularity at the origin in the relativistic model, we modified this to include a maximum central density. We can also evaluate analytically the lensing properties of such a cluster in the $\Lambda$CDM model, where $\Omega_{m}^{0}=0.27$, $\Omega_{\Lambda}^{0}=0.73$, and $H_0=69$ km s$^{-1}$ Mpc$^{-1}$. The density of the cluster is isotropic and given by
\be
\rho^{\textrm{BMO}}(x)=\frac{\delta_c \rho_c}{\left(\epsilon_c+x\right)\left(1+x\right)^2}\left(\frac{\tau^2}{x^2+\tau^2}\right)^2,\label{eq:bmodensity}
\ee
where $\rho_c$ is the critical density of the universe, $x=r/r_s$ and $\tau=r_t/r_s$ are dimensionless lengths, $r_s=r_{200}/c$ is a scale radius, $c=3$ is the concentration of the cluster, $r_{200}=1.75$ Mpc is the radius at which the cluster density is 200 times the critical density, $r_t=3$ Mpc is the truncation radius, $\epsilon_c=0.1$ enforces a peak central density of $10\delta_c\rho_c$, and the characteristic overdensity is
\be
\delta_c=\frac{200}{3}\frac{c^3}{\log{(1+c)}-\frac{c}{1+c}}.
\ee

From this density profile, we can evaluate the convergence $\kappa(x)=\Sigma(x)/\Sigma_{crit}$ and shear $\gamma(x)=\bar{\kappa}(x)-\kappa(x)$, where $\Sigma(x)$ is the surface mass density, $\bar{\kappa}(x)=M_{proj}(x)/\pi r^2\Sigma_{crit}$, $M_{proj}(x)$ is the projected mass enclosed within $r$, and 
\be
\Sigma_{crit}=\frac{D_s}{4\pi D_{\ell s} D_\ell}.\label{eq:critdens}
\ee
The surface mass density is
\bea
\Sigma(x)&=&r_s\int_{-\infty}^{\infty}d\ell \rho^{\textrm{BMO}}(\sqrt{\ell^2+x^2})\\
&=&\frac{\delta_c\rho_c r_s\tau ^4}{\left(\tau ^2+1\right)^3}  \Bigg\{-\frac{2}{\left(x^2-1\right) \left(\epsilon _c-1\right){}^2} \Big[-\epsilon_c \left(\left(\tau^2-3\right) x^2+4\right)+\tau^2+x^2+5\left(1-x^2\right)\Big]F_1(x)\nonumber\\
&&-\frac{1}{\tau ^2 \left(\tau ^2+x^2\right) \left(\epsilon _c^2+\tau ^2\right)^2}\Big[-2 \epsilon _c^3 \left(4
\tau ^4+\left(3 \tau ^2-1\right) x^2\right)-\tau ^2 \epsilon _c \left(12 \tau ^4+2 \tau ^2 \left(5 x^2+2\right)+2 x^2\right)\nonumber\\
&&+\epsilon _c^2 \left(2 \tau^6+\tau ^4 \left(x^2-6\right)-6 \tau ^2 x^2+x^2\right)+\tau ^2 \left(4 \tau ^6+3 \tau ^4 \left(x^2-2\right)-2 \tau ^2 \left(3 x^2+1\right)-x^2\right)\Big]F_2\left(\frac{x}{\tau }\right)\nonumber\\
&&-\frac{\pi}{2 \left(\tau^2+x^2\right)^{3/2} \left(\epsilon_c^2+\tau^2\right)^2}\Big[\epsilon _c^3 \left(3 \tau ^4+2 \tau ^2 \left(x^2-3\right)-6 x^2-1\right)+\epsilon _c^2 \left(6 \tau ^4+2 \tau ^2
\left(2 x^2-1\right)-4 x^2\right)\nonumber\\
&&+\epsilon _c \left(5 \tau ^6+2 \tau ^4 \left(2 x^2-3\right)-3 \tau ^2 \left(2 x^2+1\right)-2 x^2\right)+2 \tau ^4 \left(5
\tau ^2+4 x^2+1\right)\Big]\nonumber\\
&&+\left(-\tau ^2-1\right) \Big[\frac{-\epsilon_c^2+\left(\epsilon _c+1\right){}^2-\tau ^2}{\left(\tau ^2+x^2\right) \left(\epsilon_c^2+\tau ^2\right)}+\frac{2}{\left(x^2-1\right) \left(\epsilon_c-1\right)}\Big]\Bigg\}\nonumber\\
&&-\frac{2\delta_c\rho_c r_s\tau ^4}{\left(\epsilon_c-1\right)^2\left(\epsilon_c^2+\tau ^2\right)^2}F_1\left(\frac{x}{\epsilon _c}\right),\label{eq:sheet}
\eea
and the projected mass density is then given by
\bea
M_{proj}(x)&=&r_s^2\int_0^x dx'2\pi x'\Sigma(x')\\
&=&\frac{2\delta_c\rho_cr_s\tau^4}{\left(\tau ^2+1\right)^3 x^2}\Bigg\{\frac{2}{\left(\epsilon_c-1\right)^2}\Big[\epsilon_c\left(\tau^2\left(x^2-2\right)-3 x^2+2\right)+\tau^2+4 x^2-3\Big]F_1(x)\nonumber\\
&&-\frac{1}{\tau^2\left(\epsilon_c^2+\tau^2\right)^2}\Big[-2\tau^2\epsilon_c\left(4\tau^4+\left(5\tau^2+1\right)x^2\right)+2\epsilon_c^3 \left(-2 \tau ^4+\tau^2 \left(2-3 x^2\right)+x^2\right)\nonumber\\
&&+\epsilon_c^2 \left(\tau^4 \left(x^2-6\right)+2 \tau^2 \left(1-3 x^2\right)+x^2\right)+2\tau^8+3 \tau^6 \left(x^2-2\right)-6 \tau^4 x^2-\tau^2x^2\Big]F_2\left(\frac{x}{\tau}\right)\nonumber\\
&&-\frac{\pi}{2 \sqrt{\tau^2+x^2}\left(\epsilon_c^2+\tau^2\right)^2} \Big[\epsilon_c^3\left(\tau^4+2 \tau ^2 \left(x^2-3\right)-6 x^2+1\right)+2 \epsilon _c^2 \left(\tau ^4+\tau ^2 \left(2 x^2-3\right)-2 x^2\right)\nonumber\\
&&+\epsilon_c \left(3 \tau^6+2 \tau^4 \left(2 x^2-3\right)-\tau^2 \left(6 x^2+1\right)-2 x^2\right)+2 \tau^4 \left(3 \tau^2+4 x^2-1\right)\Big]\nonumber\\
&&+\frac{\pi}{2 \tau\left(\epsilon_c^2+\tau^2\right)^2}\Big[2 \left(\tau^2-3\right) \tau^2 \epsilon_c^2+\left(3 \tau ^4-6 \tau ^2-1\right) \tau^2 \epsilon_c+\left(\tau^4-6 \tau^2+1\right) \epsilon_c^3+6 \tau^6-2 \tau^4\Big]\nonumber\\
&&+\frac{2 }{\left(\epsilon_c^2+\tau^2\right)^2}\Big[-\tau^4 \left(4 \epsilon_c+3\right)+2 \left(1-\tau^2\right) \epsilon_c^3+\left(1-3 \tau^2\right) \epsilon_c^2+\tau^6\Big]\log(\tau )\Bigg\}\nonumber\\
&&+\frac{4\delta_c\rho_cr_s\tau^4}{\left(\epsilon_c-1\right)^2\left(\epsilon_c^2+\tau^2\right)^2x^2}\Bigg\{\left(\epsilon_c^2-x^2\right)F_1\left(\frac{x}{\epsilon_c}\right)-\epsilon_c^2\log\left(\epsilon_c\right)\Bigg\}\label{eq:projm},
\eea
where
\bea
F_1(x)&=&\begin{cases}
\frac{\arctan{\sqrt{x^2-1}}}{\sqrt{x^2-1}}& x>1 \\
\frac{\textrm{arctanh}{\sqrt{1-x^2}}}{\sqrt{1-x^2}}& x\le1
\end{cases}\\
F_2(x)&=&\frac{\textrm{arccoth}{\sqrt{1+x^2}}}{\sqrt{1+x^2}}.
\eea
In the limit that $\epsilon_c\rightarrow 0$, this results in the lensing properties found for the profile of \cite{bmo}. Also letting $\tau\rightarrow \infty$, these properties agree with the classic NFW results. The resulting convergence and shear is shown in Fig. \ref{fig:lcdmlens} for the cluster profile both with \cite{troxel1} and without the central peak density for a source at $z=1.0$ and lens at $z=0.5$. The two agree well except for near the center of the cluster, where the convergence and shear are lower in the cluster with a limiting peak density, and in the shear, where the peak density causes a small decrease in the total shear over a large range of radii.

\begin{figure}
\includegraphics[width=\columnwidth]{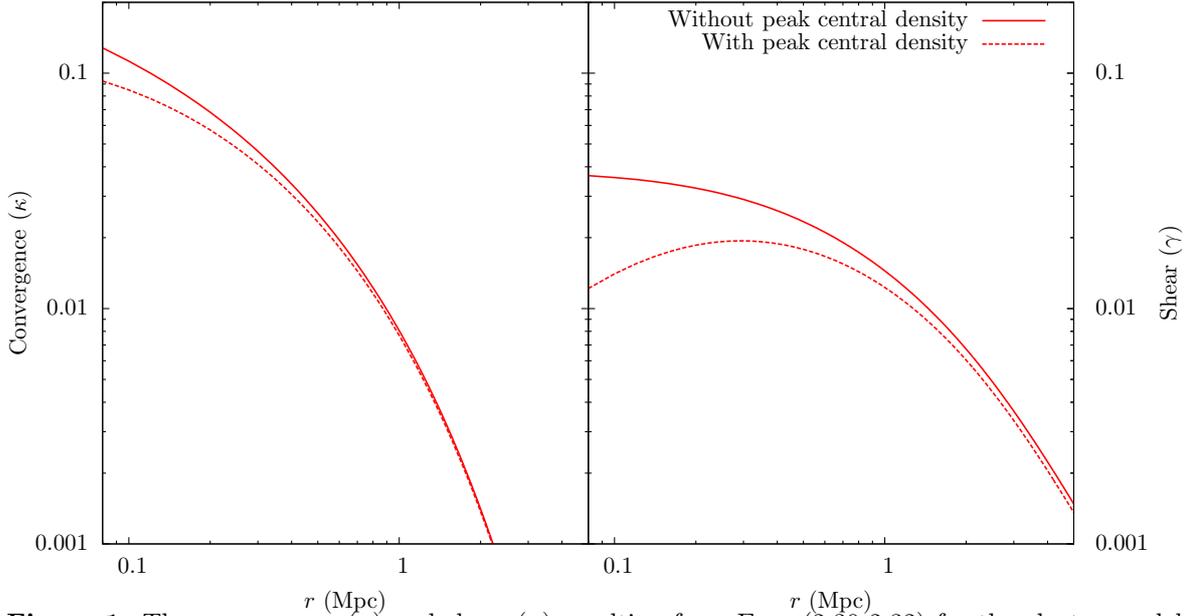}
\caption{\label{fig:lcdmlens}
The convergence ($\kappa$) and shear ($\gamma$) resulting from Eqs. (\ref{eq:sheet}-\ref{eq:projm}) for the cluster model described by the truncated NFW density profile in equation (\ref{eq:bmodensity}) is shown compared to that for the same profile with no central peak density as given in \cite{bmo}. The convergence and shear are evaluated for a source at redshift $z=1.0$ and lens at $z=0.5$. As expected, the limiting central peak density in our profile causes a reduction in the magnitude of both $\kappa$ and $\gamma$ near the center of the structure, as well as a small decrease in the total shear which extends out to large radii.}
\end{figure}

\section{Lensing properties in exact, anistropic models}\label{model}
\subsection{Anisotropic model construction}\label{model2}

The example cluster profile given in section \ref{lcdmcluster} was spherically symmetric and assumed to reside within a background FLRW spacetime. Because the derivation assumes a thin lens and utilizes the resulting surface mass densities, the background cosmology only impacts the calculation as part of deriving the angular diameter distances used in the critical surface density in equation (\ref{eq:critdens}), which are calculated ignoring the presence of the structure. Relaxing these assumptions is necessary in order to explore the degree to which introducing anisotropic structure into lensing calculations will impact the resulting convergence or shear, and whether there exist potential biases that this might introduce into our conclusions about cosmological or astrophysical quantities in our models.

Here we explore this by defining both a spherically symmetric and inhomogeneous reference model for the cluster of galaxies, which is identical in its resulting density to that used in the $\Lambda$CDM calculation in section \ref{lcdmcluster}, as well as an anisotropic model which allows us to control the degree of anisotropy. The impact of including anisotropies can then be quantified by utilizing the process described in section \ref{lensing} for both models and comparing any discrepancies between the resulting lensing properties. To do this, we will use the Lema\^itre-Tolman (LT) \cite{Lemaitre1933,tolman1934} and Szekeres \cite{Szekeres1,Szekeres2} models, as discussed below.

\subsubsection{The inhomogeneous and anisotropic Szekeres metric}\label{szekeres}

The reference LT model and Szekeres models through which we vary the levels of anisotropy are described in detail in \cite{troxel1}, but we review them here for completeness. Both models are based on an exact metric representation of the density in the universe that includes a local overdensity with radial profile consistent with a truncated NFW galaxy cluster model. The LT model is inhomogeneous, but isotropic about a single point, while the Szekeres models relax that isotropy and are fully general (i.e., they possess no Killing vectors \cite{Bonnoretal1977}). The LT models are a natural limit to the Szekeres models, and the FLRW models can be natural limits to both. 

The Szekeres metric is an exact solution to Einstein's field equations with an irrotational dust source, and its lack of symmetries makes it ideal to model general asymmetric structure in the universe. We will utilize the LT form of the Szekeres metric, which can be written in synchronous and comoving coordinates as \cite{Hellaby1996}
\be
ds^2=-dt^2+\frac{(\Phi,_{r}-\Phi \E,_{r}/\E)^2}{\epsilon-k(r)}dr^2+\frac{\Phi^2}{\E^2}(dp^2+dq^2),\label{eq:metric}
\ee
where $,_{\alpha}$ represents partial differentiation with respect to the coordinate $\alpha$. Depending on the dependencies of the metric functions on the coordinate $r$, the Szekeres models fall into two classes. Here we use the more general Class I metric, with all metric functions having an $r$-dependence. 

The geometry of the spatial sections in the metric is governed by $k(r)$, which generally depends on $r$, with open, closed and flat sections potentially existing. The parameter $\epsilon=0,\pm 1$ then determines the geometry of the $(p,q)$ 2-surfaces. In the LT form of the metric, the entire space is foliated by 2-surfaces with a single geometry, but in more general forms of the metric, this geometry can also be a function of $r$. We will limit our discussion to the quasi-spherical case ($\epsilon=1$), since they are better studied and understood. In the quasi-spherical case, the $(p,q$) 2-surfaces are spheres with $\Phi=\Phi(t,r)$ as their areal radius.

$\Phi(t,r)$ is defined by 
\be
(\Phi,_{t})^{2}=-k+\frac{2M}{\Phi}+\frac{\Lambda}{3}\Phi^2,\label{eq:phi}
\ee
where $M=M(r)$ represents the total active gravitational mass within a sphere of constant $r$. We will choose $\Lambda=0$ for simplicity, which lets us write the solution of equation (\ref{eq:phi}) in a simple parametric form. Equation (\ref{eq:phi}) has the same form as the Friedmann equation, but where each surface of constant $r$ evolves independently of the others. The function $\E=\E(r,p,q)$ in equation (\ref{eq:metric}) is
\be	
\E(r,p,q)=\frac{S(r)}{2}\left[\left(\frac{p-P(r)}{S(r)}\right)^2+\left(\frac{q-Q(r)}{S(r)}\right)^2+\epsilon\right],\label{eq:E}
\ee
and $S$, $P$, and $Q$ describe the stereographic projection from the $p$ and $q$ plane onto the unit sphere such that
\be	
\frac{(p-P(r),q-Q(r))}{S(r)}=\frac{(\cos(\phi),\sin(\phi))}{\tan(\theta/2)}.\label{eq:transform}
\ee
When $\E=\E(p,q)$ ($S$, $P$, and $Q$ are constants), equation (\ref{eq:metric}) is just the LT metric, and the spheres of constant $t$ and $r$ are concentric about the origin. The contribution of $\E,_{r}/\E$ to the Szekeres metric acts to offset these spheres relative to the LT model through the $r$ dependence of $S$, $P$, and $Q$.

The density in the Szekeres metric is given by
\be	
\kappa\rho(t,r,p,q)=\frac{2(M,_{r}-3 M\E,_{r}/\E)}{\Phi^2(\Phi,_{r}-\Phi\E,_{r}/\E)},\label{eq:density}
\ee
where $\kappa=8\pi$ in units where $c=G=1$. In an isotropic structure $\E,_{r}=0$, and the density depends only on the total mass function $M(r)$ and areal radius $\Phi(t,r)$ (and thus the coordinates $t$ \& $r$). In this case, the Szekeres and LT densities are identical. Indeed at any radius where $\E,_{r}=0$, this is true and the density along that 2-sphere is homogeneous. Substructure in the form of anisotropic perturbations on this corresponding LT structure are then due to the influence of $\E,_{r}/\E\ne0$, which introduces a dipole contribution to the density at a given $r$ \cite{Szekeres2}. This means that along any surface of constant $r$ where $\E,_{r}/\E\ne0$, there exists a single density peak which decreases monotonically to a density minimum located on the opposite side of the 2-sphere.

\subsubsection{Model properties}\label{modeldef}

Our cluster model is fully defined by the Szekeres metric functions, which have been chosen to reproduce at $t_0$ the density profile of a truncated NFW cluster that is set within a background FLRW model with $\Omega_m^0=1$. The resulting density profile, which extends beyond the cluster, then has the form
\be
\rho^{\textrm{LT}}(r)=\rho^{\textrm{BMO}}(r)+\rho_c.\label{eq:ltdensity}
\ee
In addition to the functions $M(r)$ and $k(r)$, there exists a final free function, the Bang time $t_B(r)$, which results from integrating equation (\ref{eq:phi})
\be
t-t_B(r)=\int_0^{\Phi}\frac{d\Phi'}{\sqrt{-k(r)+\frac{2 M(r)}{\Phi'}}}.\label{eq:phiint}
\ee
The function $t_B(r)$ gives the local time of the Big Bang at some $r$. Though the function $t_B(r)$ is arbitrary in general, a homogeneous model requires that $t_B(r)=\textrm{const}$.

The process to construct the necessary metric functions is discussed in \cite{troxel1}, which we will not reproduce here, since we are not modifying the models used. The resulting functions $M(r)$, $k(r)$, and $t_B(r)$ are shown in \cite{troxel1}. The curvature $k$ is everywhere positive, and at $r>50$ Mpc $k\propto r^2$ and $M\propto r^3$ as required in a homogeneous (FLRW) space. The Bang time function $t_B$ is chosen such that the structure is near the middle of its collapsing phase, where it has not yet collapsed to a point where pressure or rotation should be a dominant component of the kinematics of the cluster. At large $r$, $t_B$ becomes constant as part of ensuring that the metric becomes homogeneous outside the cluster.

The isotropic density profile in equation (\ref{eq:bmodensity}) we use for the cluster is based on a truncated version of the NFW profile. The classic NFW profile has two undesirable features: 1) its density diverges at small $r$, which will violate the conditions of regularity at the origin necessary for the Szekeres metric; 2) its enclosed mass diverges at large $r$, which makes matching to an FLRW space with homogeneous density difficult. To resolve these problems, we introduced a maximum density at very small $r$, and a truncation radius $r_t$ following \cite{bmo}. The constant density component of equation (\ref{eq:ltdensity}) is included so that the model can take on the appropriate density for an $\Omega_m=1$ FLRW model at large $r$.

\begin{figure}
\center
\includegraphics[width=0.6\columnwidth]{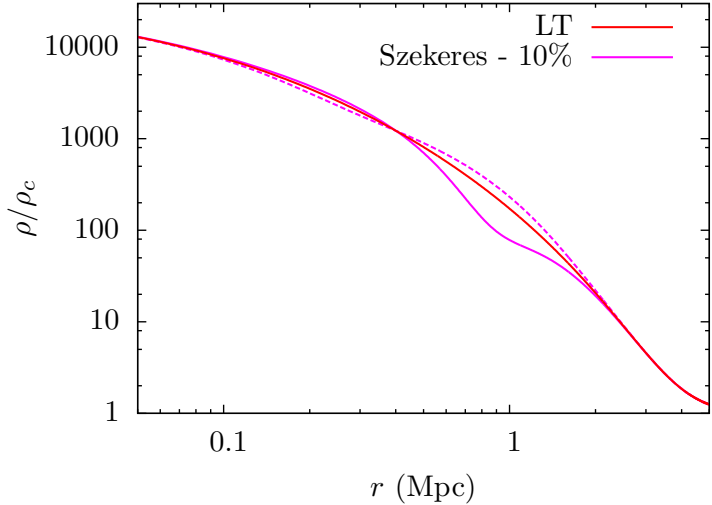}
\caption{\label{fig:rho}
The density profile of the $\delta\rho_{|\mathcal{D}}/\rho^{\textrm{LT}}_{|\mathcal{D}}=10\%$ Szekeres anisotropic model is compared to the isotropic LT spherically symmetric model. The total mass at each $r$ is identical in both models. The Szekeres curves are measured through the directions of maximum $|\rho-\rho^{\textrm{LT}}|$. Solid lines look through the $\phi=\pi/2$ direction and dashed lines look through the $\phi=-\pi/2$ direction. In both cases, $\theta=\pi/2$.}
\end{figure}

\subsubsection{Introducing anisotropic substructure}\label{anisotropic}

To introduce substructure into the galaxy cluster, and thus cause the density profile to become anisotropic, we choose a specific form for the functions $S(r)$, $P(r)$, and $Q(r)$ which compose the metric function $\E(r,p,q)$. This function controls the anisotropy in the structure, but must also satisfy several physical requirements as discussed in \cite{troxel1} in order to avoid the formation of singularities during the evolution of the structure. We will also enforce an FLRW background by requiring that $S,_{r}=P,_{r}=Q,_{r}=0$ at large $r$. Once these requirements are addressed, we define the functions as $S(r)=\textrm{const.}$, $P(r)=0$, and
\be
Q(r)=27 e^{-5r}r^2.\label{eq:q}
\ee
Varying the value of $S$ allows us to control the strength of the anisotropy. This choice of functions produces a pair of overdense/underdense dipoles in the structure's density at different values of $r$, which is shown relative to the LT density in Fig. \ref{fig:rho}. The boundary between these two anisotropic regions occurs at $r=0.4$ Mpc, where $Q,_{r}=0$. Since $S$ and $P$ are constant, this particular 2-sphere is entirely LT-like (spherically symmetric) and has a constant density. 


The strength of the anisotropy is defined as in \cite{troxel1}. This is expressed as the total fractional displaced mass present in the anisotropies at $t_0$, $\delta\rho_{|\mathcal{D}}/\rho^{\textrm{LT}}_{|\mathcal{D}}$, where $\delta\rho_{|\mathcal{D}}$ is the total displaced mass of the Szekeres structure relative to the LT reference density and $\rho^{\textrm{LT}}_{|\mathcal{D}}$ is the total LT density bounded by a 2-sphere at $r=2 r_{200}$. A discussion of the volume integral, $X_{|\mathcal{D}}$, of a quantity $X$ over some domain $\mathcal{D}$ in the Szekeres metric is given in appendix A of \cite{troxel1}.

We choose anisotropies that represent realistic estimates of the amount of the halo mass likely to be present in substructure \cite{Tormen1998}, with $\delta\rho_{|\mathcal{D}}/\rho^{\textrm{LT}}_{|\mathcal{D}}$ of $5\%$, $10\%$, and $15\%$. These anisotropies correspond to $S=4.87$, $2.43$, and $1.65$ Mpc, respectively.

Figure \ref{fig:rho} shows the resulting density anisotropies produced in the Szekeres model through the angles of largest deviation in the dipoles from the spherically symmetric LT model. We consider these anisotropies to be conservative, given that we are simply shifting mass in the underlying isotropic NFW halo with the Szekeres dipole and not adding the substructure mass to the original density.

\subsection{The geodesic deviation in the inhomogeneous and anisotropic Szekeres models}\label{geodev}

The geodesic deviation of a null ray was presented and discussed in section \ref{lensing} for a general cosmology. However, we must now specialize this framework to the Szekeres metric. In order to calculate the geodesic deviation, we must first define the components of the optical tidal matrix in equation (\ref{eq:tidaleq2}). The most important components of the optical tidal matrix are the Ricci and Weyl curvatures of the Szekeres spacetime. The Ricci curvature tensor can be expressed in terms of the Reimann curvatur tensor or Christoffel symbols as 
\be
R_{\mu\nu}=R^{\alpha}{}_{\mu\alpha\nu}=2\Gamma^{\alpha}_{\mu[\nu,\alpha]}+2\Gamma^{\alpha}_{\beta[\alpha}\Gamma^{\beta}_{\nu]\mu},\label{eq:ricci}
\ee
and the Weyl curvature tensor as  
\be
C_{\alpha\beta\mu\nu}=R_{\alpha\beta\mu\nu}-\left(g_{\alpha[\mu}R_{\nu]\beta}-g_{\beta[\mu}R_{\nu]\alpha}\right)+\frac{1}{3}R g_{\alpha[\mu}g_{\nu]\beta},\label{eq:weyl}
\ee
where $R=R^{\alpha}{}_{\alpha}$ is the Ricci scalar, $g_{\alpha\beta}$ is the metric tensor, and square brackets denote the antisymmetric part. The Christoffel symbols and the resulting Ricci and Weyl curvatures for the Szekeres metric are given in appendix A.

In addition to the curvatures listed above, the optical tidal matrix also depends on both the null tangent vector $k^{\mu}$ and the screen basis vectors $E_1^{\mu}$ and $E_2^{\mu}$. The null tangent vector can be computed from the null geodesic equation $(k^{\mu})\dot{}+\Gamma^{\mu}_{\alpha\beta}k^{\alpha}k^{\beta}=0$, which produces the set of equations \cite{Nwankwoetal2011}
\bea
0&=&\dot{k^{t}}+HH,_{t}(k^r)^2+FF,_{t}[(k^p)^2+(k^q)^2]\\
0&=&H^2\dot{k^{r}}+H\dot{H}k^r-HH,_{r}(k^r)^2-FF,_{r}[(k^p)^2+(k^q)^2]\\
0&=&F^2\dot{k^{p}}+F\dot{F}k^r-HH,_{p}(k^r)^2-FF,_{p}[(k^p)^2+(k^q)^2]\\
0&=&F^2\dot{k^{q}}+F\dot{F}k^r-HH,_{q}(k^r)^2-FF,_{q}[(k^p)^2+(k^q)^2],\label{eq:null}
\eea
where we have defined $H=(\Phi,_{r}-\Phi \E,_{r}/\E)/\sqrt{\epsilon-k(r)}$ and $F=\Phi/\E$, and $\dot{}$ represents differentiation w.r.t. the affine parameter $\lambda$. In the LT limit, $H=\Phi,_{r}/\sqrt{\epsilon-k(r)}$.

Finally, the screen basis vectors $E_1^{\mu}$ and $E_2^{\mu}$ are defined at the observer in terms of the null vector $k^{\mu}_0=(-1,k^r_0,k^p_0,k^q_0)$, the comoving velocity of the observer $u^{\mu}_0=(1,0,0,0)$, and the metric functions. $E_1^{\mu}$ and $E_2^{\mu}$ obey the following orthogonality conditions
\bea
0&=&k_{\mu}E_a^{\mu}\\
0&=&u_{\mu}E_a^{\mu}\\
\delta_{ab}&=&E_a{}_{\mu}E_b^{\mu},\label{eq:ortho}
\eea
where $a\in \{1,2\}$. These conditions result in the initial conditions
\bea
E_1^{\mu}&=&\left(0,\frac{F}{H}\sqrt{(k^p_0)^2+(k^q_0)^2},-\frac{H}{F}\frac{k^r_0 k^p_0}{\sqrt{(k^p_0)^2+(k^q_0)^2}},-\frac{H}{F}\frac{k^r_0 k^q_0}{\sqrt{(k^p_0)^2+(k^q_0)^2}}\right)\\
E_2^{\mu}&=&\left(0,0,\frac{1}{F}\frac{k^p_0}{\sqrt{(k^p_0)^2+(k^q_0)^2}},-\frac{1}{F}\frac{k^q_0}{\sqrt{(k^p_0)^2+(k^q_0)^2}}\right),\label{eq:inite}
\eea
for nonzero $k^p_0$ and $k^q_0$ (null vectors that are not initially radial). In the appropriate LT limit, this agrees with the previous results of \cite{ghassemi2009}. Once $E_1^{\mu}$ and $E_2^{\mu}$ are known at the observer, they are parallely propagated along with $u^{\mu}$ and $k^{\mu}$ along the path of the fiducial ray being considered. Thus, at other positions along the path parameterized by $\lambda$, we must similarly solve the differential equation $(E_a^{\mu})\dot{}+\Gamma^{\mu}_{\alpha\beta}E_a^{\alpha}k^{\beta}=0$ for each of $E_1^{\mu}$ and $E_2^{\mu}$. This results in a set of equations which are similar in form to equations (\ref{eq:null}).

\begin{figure}
\includegraphics[width=\columnwidth]{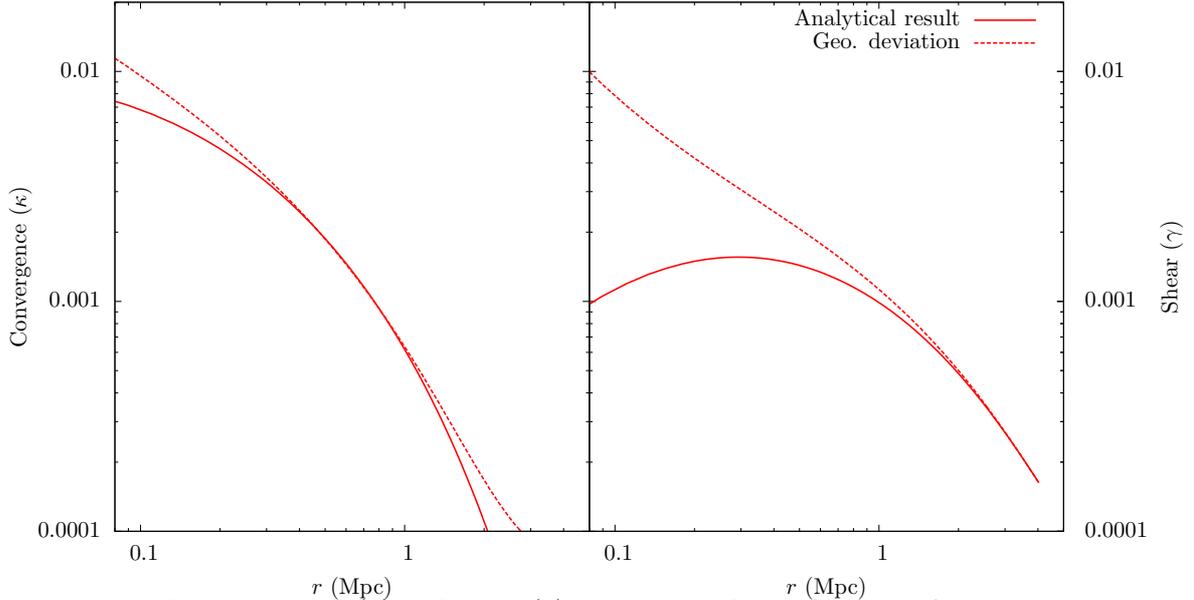}
\caption{\label{fig:nfwcomp}
The convergence ($\kappa$) and shear ($\gamma$) are compared as a function of impact parameter $r$ for the classic analytical result of a truncated NFW profile with peak central density as described in section \ref{lcdmcluster} and that found through the geodesic deviation of a ray passing through an identical and evolving structure as described in section \ref{geodev}. In both cases, the source is located at $D_s\approx 100$ Mpc and the lens is at $D_(\ell)\approx 50$ Mpc. The two methods diverge at small $r$, which is due in the second case to the structure evolving as the ray passes through it. For example, the shear should analytically go to zero at the center of a structure with a fixed, peak central density within some $r$, but will not do so in a structure that evolves, even though the density of the structure is nearly identical when the ray passes the center at approximately $t_0$. At larger $r$, the shear and convergence are consistent, though inaccuracy in the determination of the equivalent FLRW angular diameter distance used to calculate $\kappa$ and $\gamma$ in equations (\ref{eq:dtokappa}) causes the convergence to again deviate when it becomes very small. The shear does not suffer from this, as it is less sensitive to errors in $\tilde{D}_A$.}
\end{figure}

In general, equation (\ref{eq:tidaleq2}) has no analytic solution. In order to numerically evaluate it, we must instead simultaneously solve the second order differential equations for $\boldsymbol{\mathcal{D}}$ and $k^{\mu}$, along with the first order differential equations for $E_1^{\mu}$ and $E_2^{\mu}$. In sum, this requires the simultaneous numerical evaluation of 24 first order differential equations. To integrate this system, we utilize a modified fifth-order Runge-Kutta algorithm with adaptive step-size \cite{numrec}. The solution follows the propagation of the null ray past the center of the structure from a position within the exterior FLRW region ($r>50$ Mpc). Beyond this particular example of its use, it is also important to note that the evaluation of the Jacobi matrix $\boldsymbol{\mathcal{D}}$ is an alternative method to calculate the angular diameter distance in such general models, without the need to evaluate partial derivatives of the null vector.

As a verification that this process successfully reproduces the expected lensing profile for the truncated NFW profile, we evaluate the geodesic deviation for the LT structure with density profile described in equation (\ref{eq:ltdensity}) and scale it by the equivalent FLRW angular diameter distance to compare to the classic analytic expressions for the convergence and shear given by equations (\ref{eq:sheet}) \& (\ref{eq:projm}). The results of both methods are shown in figure \ref{fig:nfwcomp}, where we have a lens at $D_{\ell}\approx 50$ Mpc and source at $D_s\approx 100$ Mpc. At small $r$, both the convergence and shear are less in the analytical model, which is consistent with the reduction in shear and convergence in the profile with peak central density. This is due to density evolving as the ray passes through the cluster in the case of the numerical calculation, combined with the ray passing the center of the structure only approximately at $t_0$. In the evolving model, the density continues to grow in the center of the structure, and surpasses this peak density, which causes an amplification of the convergence and shear relative to the analytic model. In the analytical case, of course, the shear goes to zero near the center of the structure where the average enclosed mass is equal to the mass at some $r$. 

However, it is clear that the numerical calculation is successful due to the close agreement between the two shear values at larger $r$, and the agreement in the values of the convergence. The deviation at large $r$ in the convergence is due to a small systematic error in choosing the appropriate equivalent FLRW distance, which the convergence is more sensitive to at very small values (large $r$). These results lend confidence to the method, however, in agreement with previous work with the LT metric which indicate good agreement between classical lensing results and those due to exact numerical calculations of the lensing properties of spherically symmetric structures \cite{parsimood2013a,parsimood2013b}. Any systematic in the rescaling of the values due to the choice of FLRW distance will also not impact the results of this investigation, where we seek to examine the relative differences due to including anisotropy and are less interested in the absolute values of the convergence.

\subsection{The effects of structure anisotropy on lensing properties}\label{anidev}

To quantify impacts in the lensing properties due to the inclusion of anisotropies, we repeat the calculations described in the previous section for varying values of $S$. This produces anisotropies with $\delta\rho_{|\mathcal{D}}/\rho^{\textrm{LT}}_{|\mathcal{D}}$ of $5\%$, $10\%$, and $15\%$ for $S=4.87$, $2.43$, and $1.65$. To simplify discussion of the geodesic deviation, we will define several quantities in analogy to the lensing convergence and shear, 
\bea
D_{\kappa}&=&\frac{D_{22}-D_{11}}{2D_0}\\
D_{\gamma_1}&=&\frac{D_{11}+D_{22}}{2D_0}\\
D_{\gamma_2}&=&\frac{D_{12}}{2D_0}=\frac{D_{21}}{2D_0}\\
D_{\gamma}&=&\sqrt{D_{\gamma_1}^2+D_{\gamma_2}^2}.
\eea
Since we are interested primarily in relative differences, we simplify the discussion by choosing a single $D_0=100$ Mpc to be approximately the magnitude of the FLRW $\tilde{D}_A$, which is used to make the quantities dimensionless. The numerical Szekeres results for these quantities are in fact noisy due to limitations in the numerical accuracy of the integration along rays which pass through the structure at different impact parameters, and thus are smoothed using a Bezier algorithm to improve legibility of the figures that follow.

\begin{figure}
\includegraphics[width=\columnwidth]{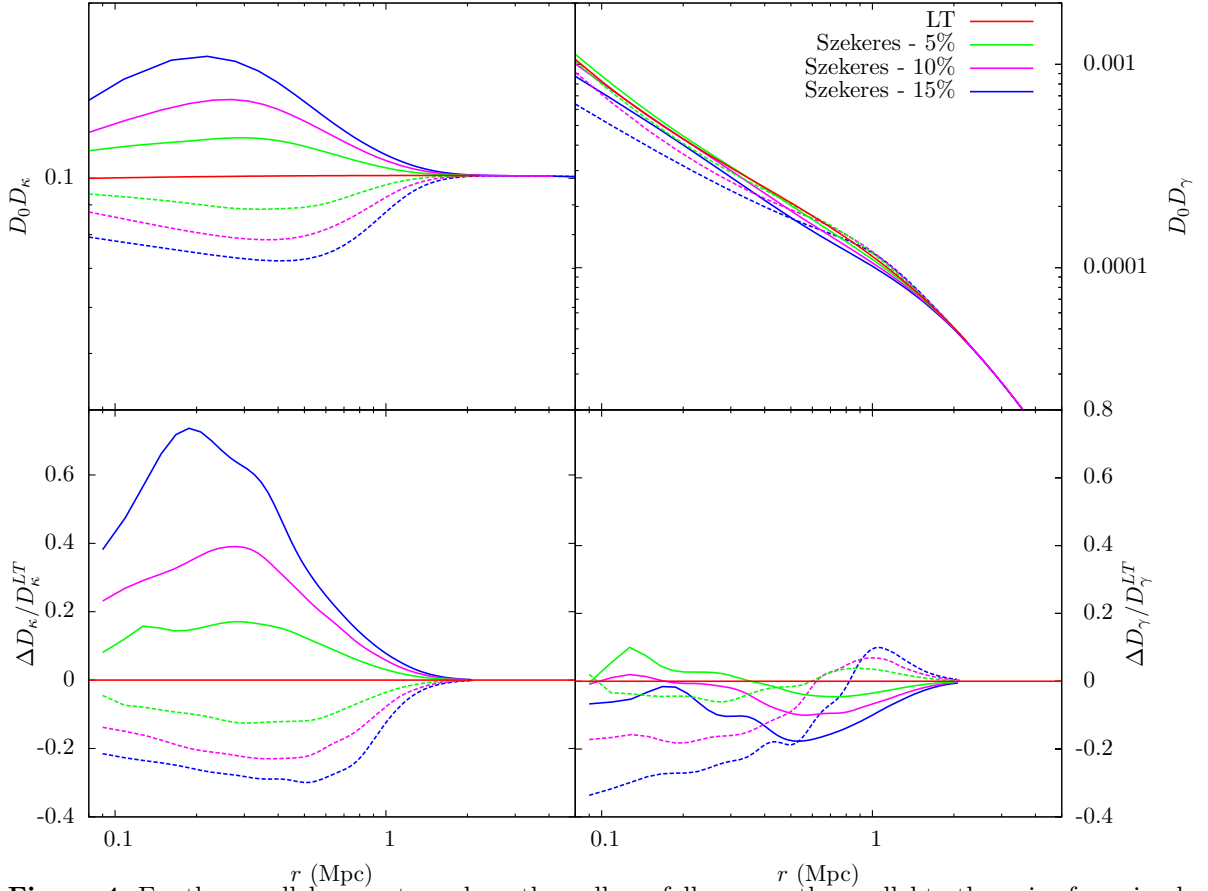}
\caption{\label{fig:cross}
For the parallel geometry, where the null ray follows a path parallel to the axis of maximal dipole anisotropy in the structure, the convergence $D_{\kappa}\approx 1-\kappa$ (left panels) and shear $D_{\gamma}\approx \gamma$ (right panels) components of the geodesic deviation are shown. Three levels of anisotropy in the Szekeres model, defined as $\delta\rho_{|\mathcal{D}}/\rho^{\textrm{LT}}_{|\mathcal{D}}$ of $5\%$, $10\%$, and $15\%$, are compared to the spherically symmetric LT reference model. In both cases there are significant deviation from the spherically symmetric (LT) case, which persists even assuming observations would be averaged over a large sample of similar, randomly aligned clusters. Top panels show deviations in magnitude while bottom panels show the relative deviations compared to the LT reference model. Solid lines represent null rays with a source at $\theta=\pi/2$ and $\phi=\pi/2$, while dotted lines have source at $\phi=-\pi/2$.}
\end{figure}

\begin{figure}
\includegraphics[width=\columnwidth]{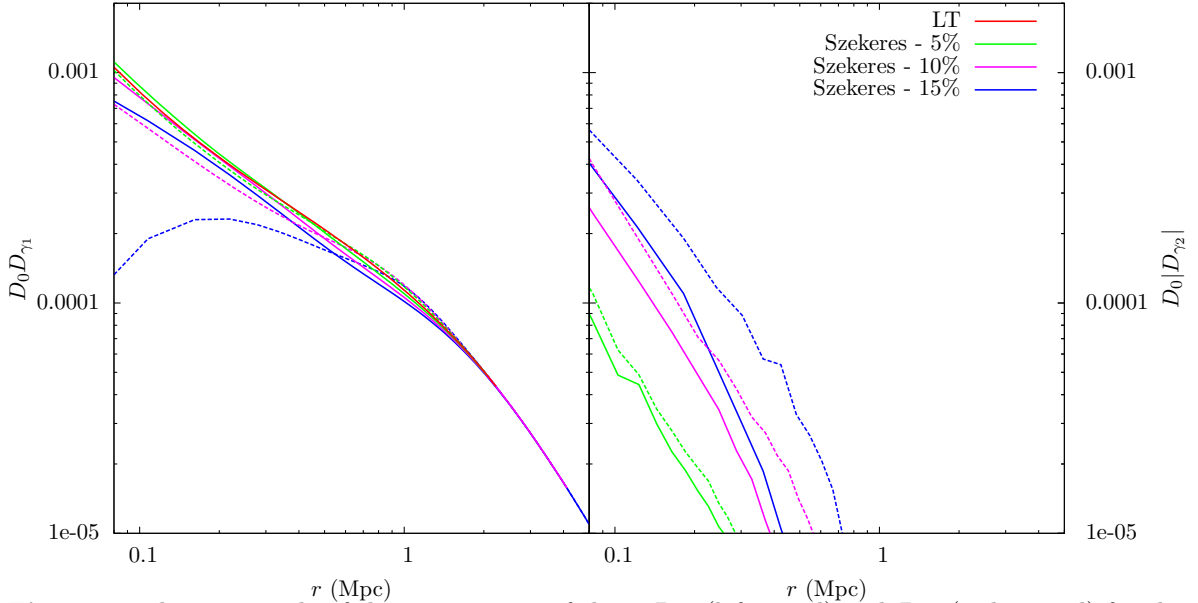}
\caption{\label{fig:crossg}
The magnitude of the components of shear $D_{\gamma_1}$ (left panel) and $D_{\gamma_2}$ (right panel) for the three anisotropic models in the parallel geometry. $D_{\gamma_2}$ is zero for the LT model, but $D_{\gamma_1}$ of the LT model is shown for comparison. In each case, $D_{\gamma_2}$ is negative for a source at $\phi=\pi/2$ and positive for a source at $\phi=-\pi/2$, while $D_{\gamma_1}$ is always positive. The $D_{\gamma_1}$ component is dominant for all models but the $15\%$ anisotropic model with source at $\phi=-\pi/2$, where $D_{\gamma_2}$ is largest at small $r$. Solid lines represent null rays with a source at $\theta=\pi/2$ and $\phi=\pi/2$, while dotted lines have source at $\phi=-\pi/2$.}
\end{figure}

\subsubsection{Null rays propagating in directions parallel to the axis of maximum dipole anisotropy}\label{cross}

We first consider a geometry that should cause maximal effect on the deviation of neighboring geodesics, where the null ray passes through the cluster parallel to the axis of maximum Szekeres dipole. This corresponds to rays initially at angular coordinates $\theta=\pi/2$ and $\phi=\pm \pi/2$, which matches the directions for which the density profile is shown in figure \ref{fig:rho}. We first propagate the null rays through the structure at varying impact parameters $r$ and plot the resulting `convergence' $D_{\kappa}\approx 1-\kappa$ and total magnitude of `shear' $D_{\gamma}\approx \gamma$. This is shown in figure \ref{fig:cross} for both absolute and relative comparisons with the LT structure. 

It is clear that the presence of anisotropy causes a strong effect on this measure of convergence (left panels) in both directions parallel to the dipole, and that the effects are not symmetric. This second effect can be easily understood, since one dipole substructure is nearer to the center of the structure and denser. This will either increase or decrease the magnitude of convergence based on whether its positive mass contribution is on the same side of the structure as the source. The increase in $D_{\kappa}$ corresponds to when the positive half of this dipole is nearest the source. While the difference appears dramatic, the net effect of the anisotropy on a randomly aligned sample of such clusters would be a much more modest increase in $D_{\kappa}$, but still non-zero. The change from a larger increase in $D_{\kappa}$ at low impact parameter $r$ to a larger decrease at high $r$ between the two paths coincides with the locations of the density dipole peaks in the structure as a function of $r$. In regions beyond the Szekeres dipoles, the convergence again agrees well with the LT result, indicating that the effects are local and will not impact measurements of convergence outside regions of anisotropy.

The total shear (right panels), represented by $D_{\gamma}$, is also affected by the anisotropy. In this case, though, the shear is primarily diminished in both directions, though less strongly in the direction were the larger positive mass dipole contribution is on the side of the source. There may be a similar `turnover' in the net effect on $D_{\gamma}$ between low and high $r$, but it is difficult to say based on these results if the shear does have a net increase near $r=1$ Mpc. Like the convergence, the total shear also matches well the LT value beyond $r=2$ Mpc.

While the effects of anisotropy on the total shear are interesting, we can also consider the two components of shear $D_{\gamma_1}\approx \gamma_1$ (left panel) and $D_{\gamma_2}\approx\gamma_2$ (right panel) separately. These are shown in figure \ref{fig:crossg}, which compares again the three Szekeres models to the spherically symmetric LT model. In the LT model, $D_{\gamma_2}=\gamma_2=0$, and the only component of the shear is $\gamma_1$. $D_{\gamma_1}$ follows a very similar trend to $D_{\gamma}$, and tends to become less than in the LT case as $r$ decreases. The $\gamma_2$ component begins as zero in the Szekeres models for large $r$, consistent with the LT model, but takes on a large negative value as the ray passes closer to the center of the structure. In fact, the magnitude of $\gamma_2$ becomes larger than that of $\gamma_1$ for the $15\%$ anisotropic model and dominates the contribution to total shear at small $r$. In all cases, $\gamma_2$ is negative for a source at $\phi=\pi/2$ and positive for a source at $\phi=-\pi/2$. This indicates that the anisotropies are in fact causing more complex changes to the shear than simply scaling its magnitude.

\begin{figure}
\includegraphics[width=\columnwidth]{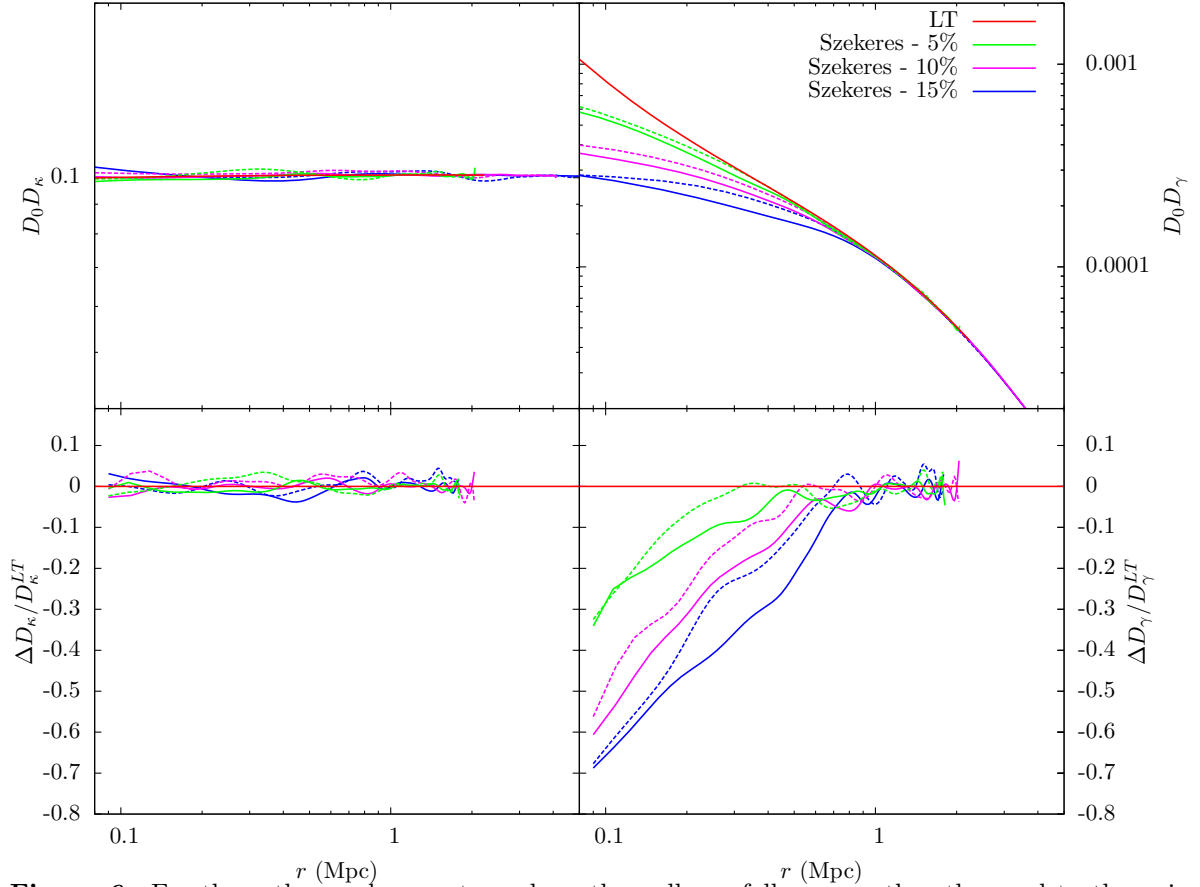}
\caption{\label{fig:orth}
For the orthogonal geometry, where the null ray follows a path orthogonal to the axis of maximal dipole anisotropy in the structure, the convergence $D_{\kappa}\approx 1-\kappa$ (left panels) and shear $D_{\gamma}\approx \gamma$ (right panels) components of the geodesic deviation are shown. Three levels of anisotropy in the Szekeres model, defined as $\delta\rho_{|\mathcal{D}}/\rho^{\textrm{LT}}_{|\mathcal{D}}$ of $5\%$, $10\%$, and $15\%$, are compared to the spherically symmetric LT reference model. The convergence shows no detectable deviation from the spherically symmetric (LT) case, while the shear is significantly reduced for sources in both directions. Top panels show deviations in magnitude while bottom panels show the relative deviations compared to the LT reference model. Solid and dotted lines represent null rays with a source at $\theta=\pi/2$ and $\phi=0,\pi$.}
\end{figure}

\begin{figure}
\includegraphics[width=\columnwidth]{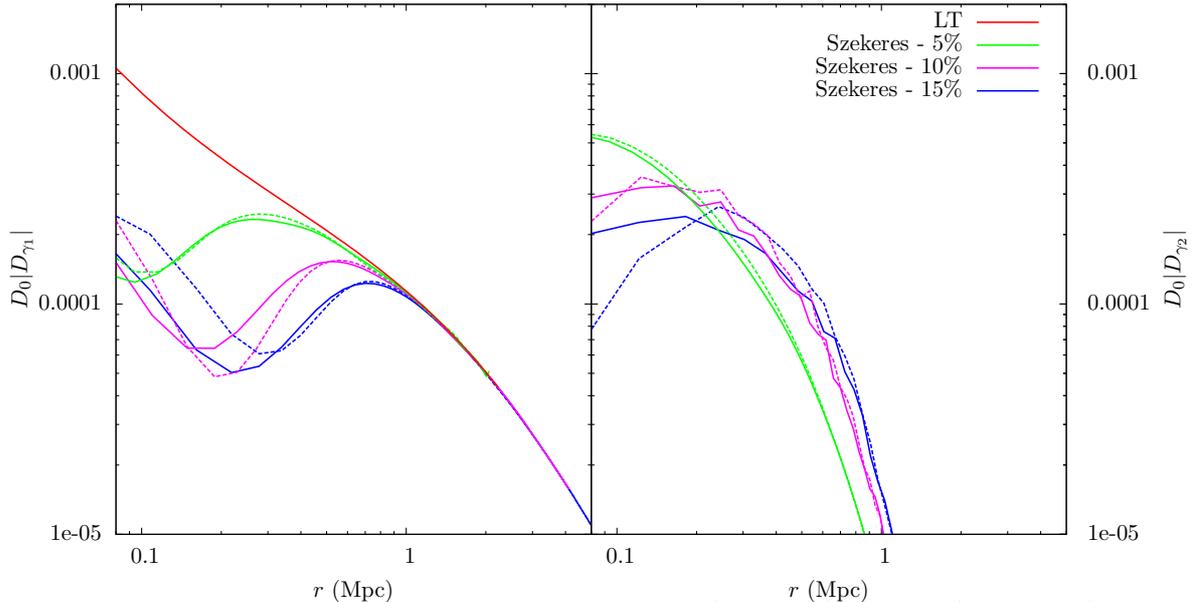}
\caption{\label{fig:orthg}
The magnitude of the components of shear $D_{\gamma_1}$ (left panel) and $D_{\gamma_2}$ (right panel) for the three anisotropic models in the orthogonal geometry. $D_{\gamma_2}$ is zero for the LT model, but $D_{\gamma_1}$ of the LT model is shown for comparison. In each case, $D_{\gamma_2}$ is positive, while $D_{\gamma_1}$ is decreases, becoming negative before decreasing in absolute magnitude again at small $r$. The $D_{\gamma_2}$ component is dominant at small $r$ for all models but the $15\%$ anisotropic model, where $D_{\gamma_1}$ grows negative enough to again dominate. Solid and dotted lines represent null rays with a source at $\theta=\pi/2$ and $\phi=0,\pi$.}\end{figure}

\subsubsection{Null rays propagating in directions orthogonal to the axis of maximum dipole anisotropy}\label{orth}

To compare to the results in section \ref{cross}, we also choose an orthogonal geometry, which one expects to cause minimal effects on the deviation of neighboring geodesics. In this geometry, the null ray passes through the cluster perpendicular to the axis of maximum Szekeres dipole and thus is exposed to the least amount of deviation in the density compared to the LT case. This corresponds to rays initially at angular coordinates $\theta=\pi/2$ and $\phi=0,\pi$. The path of the ray passing directly through the center of the structure in this geometry will thus experience the least possible change in density relative to the spherically symmetric LT reference structure. As in the case of paths parallel to the dipole axis, we first propagate the null rays through the structure at varying impact parameters $r$ and plot the resulting $D_{\kappa}\approx 1-\kappa$ and $D_{\gamma}\approx \gamma$. This is shown in figure \ref{fig:orth} for both absolute and relative comparisons with the LT structure. 

The left panels showing $D_{\kappa}$ are consistent with our picture of minimal interference in the lensing properties of the null ray along this direction. In fact, there appears to be no effect on the convergence of the ray within the accuracy of the numerical calculations, and we can rule out any effects greater than a few percent relative to the expected LT convergence. The total shear (right panels), however, indicate a much stronger impact on the shear in this geometry than in the geometry parallel to the dipole axis. It is also consistent with a reduction of the total magnitude of shear, but at much stronger levels and with better agreement between the two directions, since they are approximately symmetric in this case. However, the shear again agrees with the LT result beyond $r=0.8$ Mpc.

The resulting picture is even more interesting in this geometry for the shear when one considers the two components independently, which is shown in figure  \ref{fig:orthg} . While null rays from sources at both $\phi=0,\pi$ have positive $\gamma_2$, the magnitude grows much more strongly at low $r$ in this case and provides the dominant contribution to $\gamma$ in the $5$ and $10\%$ anisotropic cases. The magnitude of $\gamma_1$ still dominates at larger $r$, but as $r$ decreases, $\gamma_1$ is positive and decreases before becoming negative and increasing in absolute magnitude once again. This occurs at higher $r$ the larger the anisotropies in the structure, with $|\gamma_1|$ growing large enough to provide the dominant contribution to $\gamma$ again at the center of the structure.  

\begin{figure}
\includegraphics[width=\columnwidth]{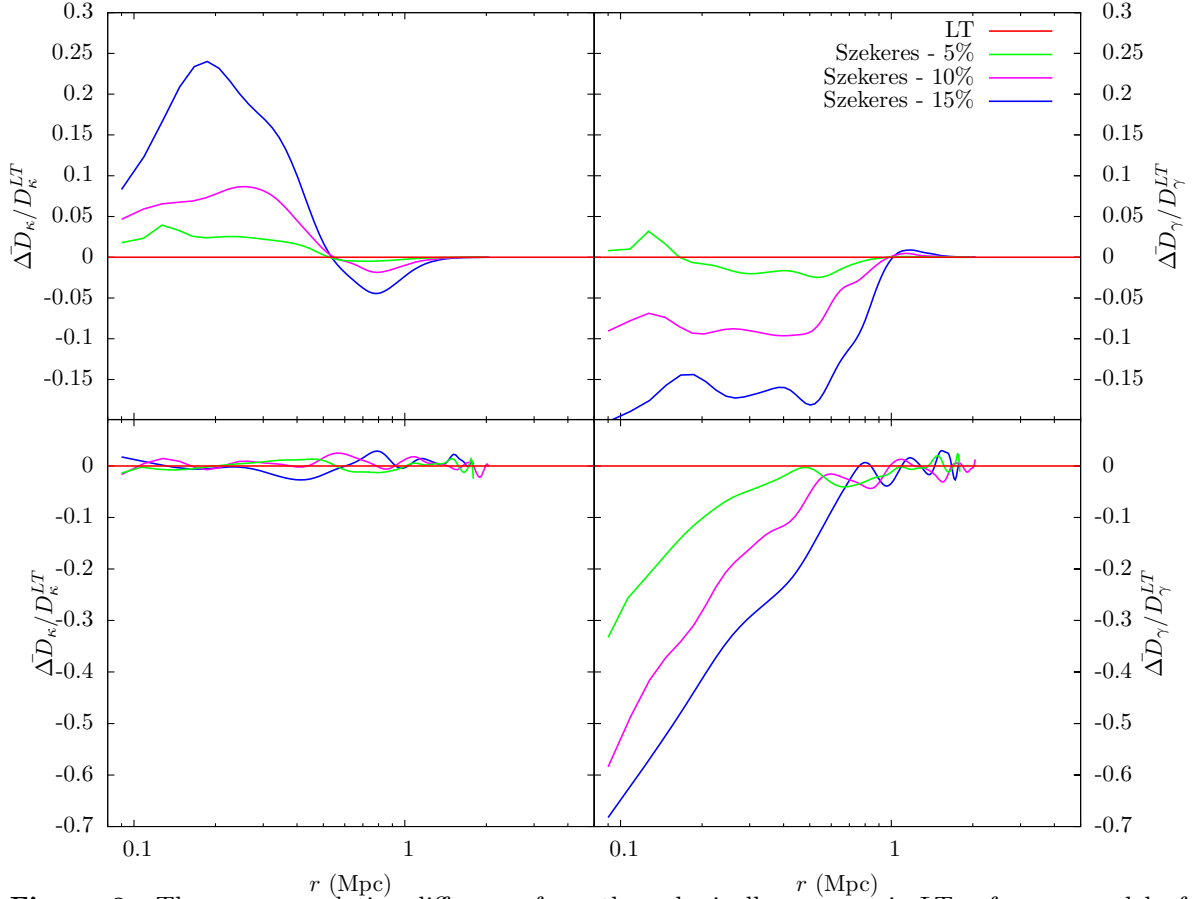}
\caption{\label{fig:sum}
The average relative difference from the spherically symmetric LT reference model of the two directions of sources, $\phi=\pm\pi/2$ or $\phi=0,\pi$, for $D_{\kappa}$ and $D_{\gamma}$ of the parallel (top panels) and orthogonal (bottom panels) geometries, respectively. This simplifies the information found in figures \ref{cross} \& \ref{orth}. The strong nonlinear dependence on the strength of anisotropy found in \cite{troxel1} is evident in the effects on the convergence for the parallel geometry, while there is no discernable impact on convergence in the orthogonal geometry. The shear tends to be less than the spherically symmetric case in both geometries, but doesn't share the particular nonlinear dependence found for the convergence or infall velocities in the models.}
\end{figure}

\subsubsection{Persistence of lensing effects in a statistical ensemble of galaxy clusters}\label{ensemble}

In sections \ref{cross} \& \ref{orth}, we explored the effects on convergence and shear of null rays which pass through specific orientations of dipole substructure anisotropy in a galaxy cluster. However, the effects depended strongly on which geometry was used, and the effects on convergence, for example, depended on which direction the dipole anisotropy was oriented relative to the source. To consider a more realistic measure of how the convergence and shear would be impacted when studying a statistical ensemble of randomly oriented galaxy clusters, we average these effects for each geometry among the two directions which the dipole anisotropy might be oriented relative to the source. This provides a more realistic and cleaner picture of the impacts on the convergence and shear measures that each geometry would have and allows us to clearly quantify the effects of varying the strength of the anisotropy.

We show these averages in figure \ref{fig:sum} for the two geometries, where the dipole is parallel (top panels) or orthogonal (bottom panels) to the propagation of the light from the source. For the parallel geometry, we see that the average impact on $D_{\kappa}$ around $r=0.2$ Mpc is an increase of about $2\%$, $8\%$, and $24\%$  for the $5\%$, $10\%$, and $15\%$ anisotropies, while at around $r=0.8$ Mpc, there is a corresponding decrease of about $1\%$, $2\%$, and $5\%$, each relative to the values obtained for a spherically symmetric LT reference structure. In the orthogonal geometry, there is of course no effect on the convergence due to anisotropy. These values for the parallel geometry are consistent with the general results we found for the impact of anisotropy in the infall velocity of dust in the same structure model, and are again very nonlinear with respect to the increase of anisotropy strength. 

For $D_{\gamma}$ in the parallel geometry, there appears to be little net effect in the $5\%$ model, though for anisotropies of $10\%$ and $15\%$, there is an average decrease in the shear of $8\%$ and $17\%$ for $r<0.5$ Mpc. In the orthogonal geometry, though, there is a net decrease beginning at $r=0.6$ Mpc that increases to about $30\%$, $53\%$, and $65\%$  for the $5\%$, $10\%$, and $15\%$ anisotropies at $r=0.1$ Mpc. The total impact in a full ensemble of clusters modeled in this way (including all possible geometries) is indicated to be primarily a systematic decrease in convergence and shear in anisotropic regions of the structure, given the results of figure \ref{fig:sum}. While utilizing a fully random selection of all possible geometries will ultimately result in a smaller impact on the convergence and shear than indicated in figure \ref{fig:sum}, it is clear from these results that a net effect will persist and will not be averaged out, due to the nonlinear and asymmetric nature of the deviations due to anisotropies. For example, when taking an average over such clusters which are aligned in various directions with their dipole axis either parallel and orthogonal to the source and observer, we find relative differences of up to $1\%$, $4\%$, and $12\%$ in the convergence, and of $15\%$, $32\%$, and $44\%$ in the magnitude of shear.

In a traditional lensing framework, where the mass is considered as a projection onto some surface orthogonal to the line of sight, this information on anisotropy is naturally lost. This occurs in two ways. First, when one projects a structure with the Szekeres dipole axis along the line of sight, there is no difference in the projected mass $\Sigma$ from a spherically symmetric structure. That is, the overdensity on one side of the structure is cancelled out by the underdensity on the opposite side. This is at odds with the real lensing results shown in Fig. \ref{fig:cross}, where there is clearly a non-negligible impact on convergence and shear measures due to the anisotropy. Second, one can consider the case of a projection of the structure where the Szekeres dipole axis is orthogonal to the line of sight. There is a residual anisotropy in the surface density $\Sigma$, but any of the nonlinearity exhibited by the true lensing results is destroyed in the traditional lensing framework. In the latter, one expects that the impact on the surface density due to the projected anisotropy is equal and opposite on either side of the structure's center in the plane of projection. Thus, any average over a large ensemble will result in a statistical lensing signal which is consistent with a spherically symmetric set of cluster lenses while it should not. 

\section{Conclusion}\label{conclusion}

Gravitational lensing has been demonstrated to be a powerful probe for studying the universe, either through strong and weak lensing by galaxies and clusters of galaxies or by the weak lensing or cosmic shear due to large-scale structure. The swiftly growing and improving sets of data we will collect over the coming decades beg to be met with ever more sophisticated models, where inhomogeneities, anisotropies, and nonlinearities are properly accounted for as present in the true lumpy universe. One way of doing this is by developing exact models in general relativity that are both inhomogeneous and anisotropic to represent realistic structure in the universe, and which are naturally nonlinear. The development of these exact models for use in comparing to cosmological and astrophysical observation is an ongoing process. The identification and development of observables like infall velocities and gravitational lensing in such general exact models is not always straightforward, but it is a necessary step in order to utilize and constrain these models. 

We have presented here a general framework for studying gravitational lensing based on the inhomogeneous and anisotropic Szekeres models, one of the best candidates of exact solutions discovered thus far for studying general structure in the universe. As a test of the framework, we used a realistic galaxy cluster model developed in \cite{troxel1}, which has a modified NFW density profile, to examine the impact of including anisotropies in the exact structure on lensing observables related to the convergence and shear. The results from this process are consistent in the spherically symmetric limit with previous results for the LT models and the convergence and shear for the same density profile in the classic lensing formalism. We found that the introduction of anisotropies in the structure modifies the usual lensing results in a significant quantitative way, as summarized below.

Specifically, we compared geometries with the paths of null rays intersecting the cluster at various impact parameters $r$ both parallel and orthogonal to the axis of maximal Szekeres dipole anisotropy. We find that for anisotropies of $5\%$, $10\%$, and $15\%$ of the total mass, there is a net reduction in the magnitude of convergence and shear through anisotropic regions in the model compared to the usual spherically symmetric case. The convergence is impacted to a large degree when the light passes parallel to the dipole, and the effect is strongly nonlinear with respect to the amount of anisotropy. This is consistent with findings in \cite{troxel1} for the infall velocity. The effect is also asymmetric with respect to the direction of the light propagation. This  causes the impact to persist even when averaged over path direction at the levels of about a $2\%$, $8\%$, and $24\%$ increase near $r=0.2$ Mpc, while at around $r=0.8$ Mpc, there is a corresponding decrease of about $1\%$, $2\%$, and $5\%$, respectively for $5\%$, $10\%$, and $15\%$ levels of anisotropy. There is no detectible impact on convergence where the light's path is orthogonal to the dipole axis.

The total magnitude of shear in both geometries is generally reduced w.r.t. the spherically symmetric case, by an average between the two geometries of $15\%$, $32\%$, and $44\%$, respectively, but more strongly when the ray passes orthogonally to the dipole axis, opposite the behavior of the convergence. The impact on shear is also strongly nonlinear, but decreases at higher levels of anisotropy instead of increasing, as with the convergence. When one considers both components of the shear, though, the results are more complex than the scaling of the total magnitude might indicate. For example, the component that is non-zero in the spherically symmetric case changes sign at low $r$ in the orthogonal geometry, while the zero component in the spherically symmetric case can become the dominant contributor to the total magnitude in either geometry at low $r$, and can also change sign depending on the direction of the light propagation.

This work represents a necessary first step to exploring gravitational lensing in more general, anisotropic exact models. Further refinement of the process is necessary, and includes improvements of numerical accuracy and stability in such sophisticated models. The work served to test the framework and indicates interesting implications for lensing by exact structure models when anisotropy is included with effects both on convergence and shear. The work constitutes the basis for future analysis of potential biases to statistical measurements of lensing by clusters and galaxies, with an initial indication that the effects on convergence and shear in anisotropic regions will persist at some level even when averaged statistically over all geometries. Similarly, additional applications of the framework to other lensing observables like the bending angle, time delays, and mass estimates are also left for a follow-up paper.

\acknowledgments

We thank L. King for useful comments during the preparation of this work. MT acknowledges that this work was supported in part by the NASA/TSGC Graduate Fellowship program. MI acknowledges that this material is based upon work supported in part by by NSF under grant AST-1109667 and by NASA under grant NNX09AJ55G, and that part of the calculations for this work have been performed on the Cosmology Computer Cluster funded by the Hoblitzelle Foundation.

\appendix

\section{The Christoffel symbols and Ricci and Weyl curvatures in the Szekeres metric}\label{app1}

In order to calculate the geodesic deviation in the Szekeres spacetime, one must parallely propagate both a fiducial null ray $k^{\mu}$ and the screen basis vectors $E_1^{\mu}$ and $E_2^{\mu}$, which, along with the comoving velocity of the observer $u^{\mu}$, form a tetrad at each point along the path of the fiducal ray affinely parameterized by $\lambda$. One must also know the Ricci and Weyl curvature of the spacetime at each point along the path. This first requires the calculation of the nonzero Christoffel symbols. For the Szekeres metric they are
\begin{alignat}{3}
\Gamma^t_{rr}&=HH,_t\qquad&&\Gamma^r_{pp}=\Gamma^r_{qq}=-\frac{FF,_r}{H^2}\qquad&&\Gamma^p_{pr}=\Gamma^p_{rp}=\frac{H,_r}{F}\\
\Gamma^t_{pp}&=FF,_t\qquad&&\Gamma^r_{rt}=\Gamma^r_{tr}=\frac{H,_t}{H}\qquad&&\Gamma^p_{pq}=\Gamma^p_{qp}=\frac{H,_q}{F}\\
\Gamma^t_{qq}&=FF,_t\qquad&&\Gamma^r_{rp}=\Gamma^r_{pr}=\frac{H,_p}{H}\qquad&&\Gamma^q_{pp}=-\Gamma^q_{qq}=-\frac{F,_q}{F}\\
\Gamma^r_{rr}&=\frac{H,_r}{H}\qquad&&\Gamma^r_{rq}=\Gamma^r_{qr}=\frac{H,_q}{H}\qquad&&\Gamma^q_{qt}=\Gamma^q_{tq}=\frac{F,_t}{F}\\
\Gamma^p_{rr}&=-\frac{HH,_p}{F^2}\qquad&&\Gamma^p_{pp}=-\Gamma^p_{qq}=\frac{F,_p}{F}\qquad&&\Gamma^q_{qr}=\Gamma^q_{rq}=\frac{H,_r}{F}\\
\Gamma^q_{rr}&=-\frac{HH,_q}{F^2}\qquad&&\Gamma^p_{pt}=\Gamma^p_{tp}=\frac{F,_t}{F}\qquad&&\Gamma^q_{qp}=\Gamma^q_{pq}=\frac{H,_p}{F}.\label{eq:chr}
\end{alignat}

Once these are known, the Ricci and Weyl curvatures can be calculated. The nonzero components of the Ricci tensor are
\bea
R_{tt}&=&-\left(\frac{H,_{tt}}{H}+2\frac{F,_{tt}}{F}\right)\\
R_{rr}&=&-\left(2\frac{F,_{rr}}{F}-2\frac{H,_{r}F,_{r}}{HF}+H^2\left[-\frac{H,_{tt}}{H}+\frac{H,_{pp}}{HF^2}+\frac{H,_{qq}}{HF^2}-2\frac{F,_{t}H,_t}{HF}\right]\right)\\
R_{pp}&=&F^2\left(\frac{F,_{tt}}{F}+\frac{F,_t^2}{F^2}+\frac{H,_tF,_t}{HF}-\frac{1}{H^2}\left[\frac{F,_{rr}}{F}+\frac{F,_r^2}{F^2}-\frac{H,_rF,_r}{HF}\right]\right)\\&-&\frac{F,_{pp}}{F}-\frac{F,_{qq}}{F}+\frac{F,_p^2}{F^2}+\frac{F,_q^2}{F^2}+\frac{H,_pF,_p}{HF}-\frac{H,_qF,_q}{HF}-\frac{H,_{pp}}{H}\nonumber\\
R_{q}&=&F^2\left(\frac{F,_{tt}}{F}+\frac{F,_t^2}{F^2}+\frac{H,_tF,_t}{HF}-\frac{1}{H^2}\left[\frac{F,_{rr}}{F}+\frac{F,_r^2}{F^2}-\frac{H,_rF,_r}{HF}\right]\right)\\&-&\frac{F,_{pp}}{F}-\frac{F,_{qq}}{F}+\frac{F,_p^2}{F^2}+\frac{F,_q^2}{F^2}-\frac{H,_pF,_p}{HF}+\frac{H,_qF,_q}{HF}-\frac{H,_{qq}}{H},\nonumber\label{eq:ricci2}
\eea
and the nonzero components of the Weyl tensor are
\begin{align}
C_{trtr}&=-\frac{H^2F,_p^2}{3F^4}-\frac{H^2F,_q^2}{3F^4}+\frac{H^2F,_{pp}}{3F^3}+\frac{H^2F,_{qq}}{3F^3}-\frac{H^2F,_t^2}{3F^2}-\frac{HH,_{pp}}{6F^2}\\&-\frac{HH,_{qq}}{6F^2}+\frac{F,_r^2}{3F^2}+\frac{H^2F,_{tt}}{3F}+\frac{H,_rF,_r}{3HF}+\frac{HH,_tF,_t}{3F}-\frac{F,_{rr}}{3F}-\frac{HH,_{tt}}{3}\nonumber\\
C_{tptp}&=\frac{1}{4HF}\left[F\left(H,_{qq}-H,_{pp}\right)+2\left(H,_pF,_p-H,_qF,_q\right)\right]-\frac{F^2}{2H^2}C_{trtr}\\
C_{tqtq}&=-\frac{1}{4HF}\left[F\left(H,_{qq}-H,_{pp}\right)+2\left(H,_pF,_p-H,_qF,_q\right)\right]-\frac{F^2}{2H^2}C_{trtr}\\
C_{rprp}&=\frac{H}{4F}\left[F\left(H,_{qq}-H,_{pp}\right)+2\left(H,_pF,_p-H,_qF,_q\right)\right]+\frac{F^2}{2}C_{trtr}\\
C_{rqrq}&=-\frac{H}{4F}\left[F\left(H,_{qq}-H,_{pp}\right)+2\left(H,_pF,_p-H,_qF,_q\right)\right]+\frac{F^2}{2}C_{trtr}\\
C_{pqpq}&=-\frac{F^4}{H^2}C_{trtr},\label{eq:weyl2}
\end{align}
with the additional components due to symmetries in the Weyl tensor being suppressed.

\end{document}